\documentclass[prd,aps,12pt]{revtex4}
\usepackage[dvips]{graphicx}
\usepackage{slashed}
\usepackage{amssymb}
\usepackage{natbib}
\usepackage{amsmath}
\usepackage{url}
\usepackage{color}
\begin{document}
\renewcommand{\baselinestretch}{1.5}
\newcommand{\Prd}{Phys. Rev D}
\newcommand{\Prl}{Phys. Rev. Lett.}
\newcommand{\Pl}{Phys. Lett.}
\newcommand{\Cqg}{Class. Quantum Grav.}
\newcommand{\Sch}{Schwarzschild$\;$}
\def \vx{\vec x}
\def \v{{\vec v}}
\def \vomega{\vec \omega}
\def \vnabla{\vec\nabla}
\def \vxi{\vec\xi}
\def\d{{\mathrm{d}}}
\def\half{\frac 1 2}
\def\a{\alpha}
\def\b{\beta}
\def\part{\partial}

\def\setR{\mathbb{R}}
\def\setN{\mathbb{N}}
\def\setC{\mathbb{C}}
\def\calH {{\cal H}}
\def\ie {{i.e.}}
\def\sgn{\mathrm{sgn}}

\def \hz{{\hat {\bf z}}}
\def \hx{{\hat {\bf x}}}
\def \hy{{\hat {\bf y}}}
\newcommand{\bu}{$\star\;$}
\newcommand{\de}{\delta}
\newcommand{\ga}{\gamma}
\newcommand{\ochi}{\overline\chi}
\newcommand{\ba}{\begin{array}}
\newcommand{\ea}{\end{array}}
\def\setR{\mathbb{R}}
\def\setN{\mathbb{N}}
\def\setC{\mathbb{C}}
\def\calH {{\cal H}}
\def\ie {{i.e.}}
\def\sgn{\mathrm{sgn}}
\newcommand{\norm}[1]{\parallel\!#1\!\parallel}
\newcommand{\ket}[1]{\mid\!#1\,\rangle}
\newcommand{\bra}[1]{\langle\,#1\!\mid}
\newcommand{\braket}[2]{\langle \, #1 \mid #2 \,\rangle}
\newcommand{\ddroi}[3]{\frac{d^{#1} {#2}}{d{#3}^{#1}}}
\newcommand{\dron}[3]{\frac{\partial^{#1} {#2}}{\partial{#3}^{#1}}}
\newcommand{\e}[1]{e^{#1}}
\newcommand{\sss}[1]{\scriptscriptstyle #1}
\newcommand{\bs}[1]{\boldsymbol{#1}}
\newcommand{\mathcalr}[1]{{\ensuremath{\mathscr{#1}}}}
\newcommand{\dd}{{\rm d}}
\newcommand{\GN}{G_{_\mathrm{N}}}
\newcommand{\lp}{\ell_{_\mathrm{Pl}}}
\newcommand{\fracc}[2]{\frac{\textstyle{#1}}{\textstyle{#2}}}


\title{Cosmology of a Heisenberg fluid}

\author {\textbf{S. Joffily} and \textbf{M. Novello}}

\affiliation{Instituto de Cosmologia Relatividade e Astrofisica (ICRA/CBPF) \\
Rua Dr. Xavier Sigaud 150, Urca 22290-180 Rio de Janeiro, RJ --
Brazil}
\date{\today}
\vspace{1.50cm}

\begin{abstract}
We consider a scenario in which the global geometry of the universe is driven by non-linear fermions obeying Heisenberg dynamics.
\end{abstract}

\maketitle

\section{Introduction}

In the standard framework of relativistic cosmology fields of different kinds have been analyzed as source of the geometry of the universe. This has been the case for scalar and vector fields (see \cite{sahni}, \cite{kolb} and \cite{narlikar} for references). However very few papers have been published concerning fermions in the universe as source of its geometry (see \cite{novelloivano}, \cite{novelloghost}). As far as we know, there has been no discussion concerning cosmology generated by (nonlinear) fermions. The purpose of the present work is to contribute to this analysis.

 In a series of papers (see \cite{heisenberg1}) Heisenberg examined a proposal regarding a complete quantum theory of fields and
elementary particles. Later on, in a different context, a similar approach was considered by Nambu and
Jona-Lasinio concerning a dynamical model of elementary particles \cite{nambu}. Since the original paper until to-day hundreds of
papers devoted to the NJL model were published \cite{volkov}.

It is not our intention here to
discuss this program. For our purpose, it is important only to
retain the original non linear equation of motion which Heisenberg
postulated for the constituents of the fundamental material blocks
of all existing matter. What we would like to retain
from Heisenberg's approach reduces exclusively to his suggestion
of a non linear equation of motion for a spinor field and examine the consequences of a framework in which this field can be taken as the fundamental
constituent that drives the cosmological scenario.

\section{Heisenberg dynamics}

We will call Heisenberg spinor (or H-field for short) a spinor field
that satisfies the non linear equation \cite{heisenberg1}
\begin{equation}
 i\gamma^{\mu}
\nabla_{\mu} \, \Psi - 2 s \, (A + iB \gamma^{5}) \, \Psi = 0
\label{domingo2}
\end{equation}
in which the constant $s$ has the dimension of (lenght)$^2$ and the
quantities $A$ and $B$ are given in terms of the Heisenberg spinor
$\Psi$ as:
\begin{equation}
A \equiv \overline{\Psi} \Psi \label{domingo3}
\end{equation}
and
\begin{equation}
B \equiv i \overline{\Psi} \gamma^{5} \Psi \label{domingo4}
\end{equation}
We use the convention as in \cite{novello} and the standard
definitions which just for completeness we quote:
\begin{displaymath}
\bar{\Psi} \equiv  \Psi^{+} \gamma^{0}.
\end{displaymath}
The $\gamma^{5}$ is hermitian and the others obeys the relation
\begin{displaymath}
\gamma_{\mu}^{+} = \gamma^{0} \gamma_{\mu} \gamma^{0}.
\end{displaymath}
This dynamics is obtained from the Lagrangian
\begin{equation}
L = L_{D} -  \mathbf{V} = \frac{i}{2} \,  \overline{\Psi} \,
\gamma^{\mu} \,
\nabla_{\mu} \, \Psi - \frac{i}{2} \, \nabla_{\mu} \overline{\Psi} \, \gamma^{\mu} \,
 \, \Psi  - \mathbf{V}
\end{equation}
The self-interacting term comes from a potential that can be
described in  terms either of a current-current or as a quartic of
spinors:
\begin{equation}
\mathbf{V} = s J_{\mu} \, J^{\mu}
\end{equation}
or, equivalently
\begin{equation}
\mathbf{V} = s ( A^{2} + B^{2} )
\end{equation}
The proof of this equivalence as well as the basis of most of the
properties needed to analyze non-linear spinors comes from the
Pauli-Kofink (PK) identities that establishes a set of tensor
relations concerning elements of the four-dimensional Clifford algebra of the $\gamma$ matrices. For any element $Q$ of this algebra the PK
relation states the validity of
\begin{equation}
(\bar{\Psi} Q \gamma_{\lambda} \Psi) \gamma^{\lambda} \Psi  =
(\bar{\Psi} Q \Psi)  \Psi  -  (\bar{\Psi} Q \gamma_{5} \Psi)
\gamma_{5} \Psi. \protect\label{H5}
\end{equation}
for $Q$ equal to $I$, $\gamma^{\mu}$, $\gamma_{5}$ and $\gamma^{\mu}
\gamma_{5}$. As a consequence of this relation we obtain two
extremely important consequences:
\begin{itemize}
 \item{The norm of the currents $J_{\mu} \equiv \overline{\Psi} \, \gamma_{\mu} \, \Psi $ and
 $I_{\mu} \equiv \overline{\Psi} \, \gamma_{\mu} \, \gamma_{5} \, \Psi $ have the same
strenght but opposite sign.}
 \item{The vectors  $J_{\mu}$ and $I_{\mu}$ are orthogonal.}
\end{itemize}
Indeed, using the PK relation we have
\begin{displaymath}
(\bar{\Psi}  \gamma_{\lambda} \Psi) \gamma^{\lambda} \Psi  =
(\bar{\Psi}  \Psi)  \Psi  -  (\bar{\Psi} \gamma_{5} \Psi) \gamma_{5}
 \Psi.
\end{displaymath}
Multiplying by $\bar{\Psi}$ and using the definitions above it
follows
\begin{equation}
J^{\mu} J_{\mu} = A^{2} + B^{2}. \protect\label{H6}
\end{equation}
We also have
\begin{displaymath}
(\bar{\Psi} \gamma_{5} \gamma_{\lambda} \Psi) \gamma^{\lambda} \Psi
= (\bar{\Psi} \gamma_{5} \Psi)  \Psi  - (\bar{\Psi} \Psi) \gamma_{5}
\Psi.
\end{displaymath}
From which it follows that the norm of $I_{\mu}$ is
\begin{equation}
I^{\mu} I_{\mu} = - A^{2} - B^{2} \protect\label{H7}
\end{equation}
and that the four-vector currents are orthogonal
\begin{equation}
I_{\mu} J^{\mu} = 0. \protect\label{H71}
\end{equation}
From these results it follows that the current $J_{\mu}$ is a
time-like vector; and the axial current is space-like. Thus,
Heisenberg potential $\mathbf{V}$ is nothing but the norm of the
four-vector current $J^{\mu}$.

\section{Symmetry}
Let us consider the $\gamma_{5}-$map
\begin{equation}
\widetilde{\Psi} = ( \cos \alpha + i \, \sin \alpha \,\gamma_{5} )
\, \Psi \label{agosto1045}
\end{equation}
It yields  for the scalars $ A $ and $ B $ the corresponding
changes:
\begin{eqnarray}
\widetilde{A}&=& \cos 2 \alpha \, A + sin 2 \alpha \, B \nonumber \\
\widetilde{B}&=& - \sin 2 \alpha \, A + cos 2 \alpha \, B
\end{eqnarray}
It will be convenient for latter use to define the associated scalar
field $ \varphi$ defined by
$$ \varphi \equiv A + i \, B. $$
This scalar field changes under the above map as a rotation of $ 2
\alpha:$
$$ \widetilde{\varphi} = e^{- 2 i \alpha} \, \varphi. $$
It follows that the Heisenberg potential is invariant under such map
$ \widetilde{\mathbf{V}} = \mathbf{V}.$ The kinematical part of
Lagrangian does not change if the parameter $\alpha$ is constant.
Thus, Heisenberg dynamics is invariant under such constant
$\gamma_{5}-$map.

\subsubsection{Plane wave solution of Heisenberg equation}

Although Heisenberg equation is non-linear it admits a solution as
a plane wave. Actually, it is well-known that any non-linear equation admits such
particular type of solution. We set
\begin{equation}
\Psi = e^{i k_{\alpha} x^{\alpha}} \, \Psi^{0}
\end{equation}
where $ \Psi^{0}$ is a constant spinor written in terms of
two-components spinors:
\[
\Psi^{0} =  \left(
\begin{array}{cc}
\varphi^{0} \\
 \eta^{0}  \nonumber \\
  \end{array}
\right). \]
The above decomposition implies that the two-components spinors
are not completely independent. They must satisfy the constraint
\begin{equation}
\eta^{0} = \left( \frac{2 \, i \, s \, B_{0} - \sigma_{i} k^{i}}{k_{0} - 2s
A_{o}} \right) \, \varphi^{0}.
\end{equation}
Compatibility requires the  "on-mass" condition
$$ k_{\mu} k^{\mu} = 4 s^{2} \, ( A_{0}^{2} + B_{0}^{2} ).$$

\subsection{Fundamental solution}

In linear Dirac dynamics a particular class of solutions (plane
waves) is characterized by the eigenstate property
$$  \nabla_{\mu} \Psi = i \, k_{\mu} \, \Psi. $$
In the nonlinear Heisenber dynamics it is possible to find solutions
that are defined by the property
\begin{equation}
\nabla_{\mu} \Psi = \left( a \, J_{\mu} + b\, I_{\mu}
\gamma^{5} \right)  \,\Psi \label{domingo6}
\end{equation}
where $a$ and $b$ are complex numbers of dimensionality
$(lenght)^{2}.$ It is immediate to prove that if $\Psi$ satisfies
this condition it satisfies automatically Heisenberg equation of
motion if $a$ and $b$ are such that $ 2 s = i \, (a - b).$

 Equation (\ref{domingo6}) is a rather strong condition that deals with the derivatives instead of the scalar structure obtained by the
contraction with $\gamma_{\mu}$ typical of Dirac or even for the
Heisenberg operators that appear in Dirac equation and in equation
(\ref{domingo2}). Prior to anything one has to examine the
compatibility of such condition which concerns all quantities that
can be constructed with such spinors. It is a remarkable result that
in order that the fundamental condition eq. (\ref{domingo6}) to be
integrable constants $a$ and $b$ must satisfy a unique constraint
given by $Re(a) - Re(b) = 0.$

Indeed, a direct calculation gives
\begin{equation}
\nabla_{\mu} J_{\nu} = (a + \overline{a}) J_{\mu} J_{\nu} + (b +
\overline{b}) I_{\mu} I_{\nu} \protect\label{H13}
\end{equation}
\begin{equation}
\nabla_{\mu} \, A =  (a + \overline{a}) \, A \, J_{\mu}  + (b -
\overline{b}) \, i B \,  I_{\mu} . \protect\label{H17}
\end{equation}
\begin{equation}
\nabla_{\mu} \, B =  (a + \overline{a}) \, B \,  J_{\mu}  + (b
- \overline{b}) \, i A  I_{\mu} . \protect\label{H18}
\end{equation}
\begin{equation}
\nabla_{\mu} I_{\nu} = (a + \overline{a})  J_{\mu} I_{\nu} + (b + \overline{b}) J_{\nu}
I_{\mu} . \protect\label{H16}
\end{equation}
Thus,
$$ [\nabla_{\mu}, \nabla_{\nu} ] \,\Psi =  \left(a \, \nabla_{[\mu} \,
J_{\nu]} +  b \, \nabla_{[\mu} \, I_{\nu]} \, \gamma^{5} \right)
\, \Psi. $$ Now, the derivative of the currents yields
$$ \nabla_{\mu} J_{\nu} -  \nabla_{\nu} J_{\mu} = ( a + \overline{a})
 [J_{\mu}, \,J_{\nu}] + ( b + \overline{b}) \,
[I_{\mu} , \,  I_{\nu}], $$ and
$$ \nabla_{\mu} I_{\nu} -  \nabla_{\nu} I_{\mu} = ( a + \overline{a}
- b - \overline{b})  [J_{\mu}  \, I_{\nu} - I_{\mu} \, J_{\nu}].
$$

It is a rather long and tedious work to show
that any combination $X$  constructed with $\Psi$ and for all
elements of the Clifford algebra, the compatibility requirement
$[\nabla_{\mu},
\nabla_{\nu}] X = 0$ is automatically fulfilled once the unique
condition of integrability
$$ a + \overline{a} = b + \overline{b}$$
is satisfied.

\subsection{Some useful relations}
The field $\Psi$ that we are dealing here is a 4-component quantity.
It is useful to write it in terms of two 2-component spinors under
the form

\[
\Psi =  \left(
\begin{array}{cc}
\varphi \\
 \eta  \nonumber \\
  \end{array}
\right). \]
 Then we obtain the following expressions:

\begin{equation}
A \equiv  \overline{\Psi} \,  \Psi = \varphi^{+} \varphi - \eta^{+}
\eta
\end{equation}

\begin{equation}
B \equiv  i \, \overline{\Psi} \gamma^{5} \Psi = i \, \left(
\varphi^{+} \eta -\eta^{+} \varphi \right)
\end{equation}
and for the current and the axial-current it follows:
\begin{eqnarray}
J_{0} &=& \varphi^{+} \varphi + \eta^{+} \eta \nonumber \\
J_{k} &=& \varphi^{+} \, \sigma_{k} \, \eta + \eta^{+} \, \sigma_{k}\, \varphi \nonumber \\
I_{0} &=& \varphi^{+} \eta + \eta^{+} \varphi \nonumber \\
I_{k} &=& \varphi^{+}\, \sigma_{k} \,  \varphi + \eta^{+}\,
\sigma_{k} \, \eta
\end{eqnarray}

\section{Heisenberg fluid}

From the standard definition of the energy-momentum tensor we obtain the expression of the
Heisenberg dynamics in the fundamental solution (\ref{domingo6}) as

$$ T_{\mu\nu} = l \, J_{\mu} \, J_{\nu} + n \, I_{\mu} \, I_{\nu} - s \, J^{2} \, g_{\mu\nu}. $$
where we have set $ a = a_{0} - i \, l$ and $ b = a_{0} - i \, n.$
 Let us define the four-velocity field as the normalized current $ v_{\mu} = J_{\mu} / \sqrt{J^{2}} $ and let us use this velocity
 field to decompose the energy-momentum tensor in its irreducible parts and set

 \begin{equation}
T_{\mu \nu}= (\rho + p) \, v_{\mu} \,v_{\nu}- p\, g_{\mu \nu}+q_{\mu (} v_{\nu)}+ \pi_{\mu \nu}
\end{equation}
It then follows that the heat flux  $ q_{\mu} $ vanishes identically and the remaining quantities are given by
$$ \rho =  ( l - s ) \, J^{2}$$
$$ p = \frac{(l+n)}{2} \, J^{2}$$
$$ \pi_{\mu\nu} = - \, \frac{n}{3} \,J_{\mu}\, J_{\nu} + n \, I_{\mu} \, I_{\nu} + \frac{n}{3} \, J^{2} \, g_{\mu\nu} $$

Before looking into the laws of conservation let us examine a consequence of our choice of the velocity fluid. Although the field has a self-interaction term, the fluid does not acquires
a self-acceleration. Indeed, from the definition of $v_{\mu}$ in terms of the current it follows
\begin{equation}
\nabla_{\beta} v_{\alpha} = 2 \, a_{0} \, \, \frac{I_{\alpha} \, I_{\beta}}{\sqrt{J^{2}}}
\label{8setembro1}
\end{equation}
Projecting on the direction of the velocity and using the orthogonality between the vector and the axial current it follows that $v^{\mu}$ has no acceleration.

\subsection{Heisenberg cosmology}

We will now show that such special fermion fluid can produce an isotropic world. We set for the geometry the standard spatially homogeneous and isotropic Friedman form

$$ ds^{2} = dt^{2} - S^{2}(t) \, d\sigma^{2}. $$
We limit our considerations here to the Euclidean section.

 Conservation of the vector and the axial currents are consequences of the Heisenberg dynamics:
$$\nabla_{\mu} J^{\mu} = 0.$$
$$\nabla_{\mu} I^{\mu} = 0.$$
Set
$$ \Psi = f(t) \, \Phi^{0}$$ where  $ \Phi^{0}$ is a constant 4-spinor and looking for solutions such that $ J_{\mu} = ( J, 0, 0, 0)$ and $ I_{\mu} = ( 0, I_{1}, I_{2}, I_{3}).$
Then conservation of the currents imply
$$ J = \frac{N}{S^{3}}$$
where $ N $ is a constant and $ I_{\mu} \, I^{\mu} = - J^{2}$ sets one real condition on the constant spinor. It then follows for the density of energy and pressure

\begin{equation}
 \rho = \frac{l + n}{2} \, J^{2}
 \label{251}
 \end{equation}

 \begin{equation}
  p = \frac{(3 \, l - n)}{6} \, J^{2}.
\label{252}
\end{equation}

In order to do not have anisotropies we set $ n = 0.$ This condition imply immediately that we are dealing with an ultra-relativistic fluid, that is, the equation of state is given by

$$ p= \rho.$$

The two conditions (conservation of the currents and the fundamental condition [\ref{domingo6}]) determines immediately the value of the scalar factor $ S(t)$ which is then given by
$$ S = S_{0} \, t^{1/3}.$$

It then remains to check if indeed this solution fulfills all conditions for the spinor field satisfying the special solution (\ref{domingo6}) that is, the vector current is time-like and has no spatial parts
$$ J_{k} =0$$
and the axial part is orthogonal to the vector current, which in this case reduces only to
$$ I_{0} = 0.$$
Using $ \Psi = f(t) \, \Phi^{0}$
where $ f $ is proportional to $t^{-1/2},$  the above conditions for the currents are fulfilled by setting $ \varphi^{+}_{0} \, \sigma_{k}\, \eta_{0} + h.c. = 0$ and $ \varphi^{+}_{0} \, \eta_{0} + h.c. = 0.$ This ends the proof that a Heisenberg fluid can generate a spatially homogeneous and isotropic universe.

Let us note the remarkable result that we have obtained a specific form for the geometry without using the equations of general relativity. In the present case these equations reduce to
$$ \frac{\dot{S}^{2}}{S^{2}} = \frac{\rho}{3}$$
$$ 2 \, \frac{\ddot{S}}{S} + \frac{\dot{S}^{2}}{S^{2}}  = - \, p.$$

It is a simple direct matter to verify that indeed if we substitute the density $ \rho$ and the pressure $ p $ by its values given by (\ref{251}) and (\ref{252}), these equations are automatically satisfied.

\section{Conclusion}

In this paper we have presented a specific solution of a universe in the framework of general relativity driven by a fluid identified with non-linear fermion obeying Heisenberg dynamics. The relationship between the density of energy and the pressure is that of an ultra-relativistic fluid in which the equation of state has the form $ p = \rho.$
Although we have given a simple example of an isotropic cosmology it is clear, from our equations, that anisotropic universes could as well be generated by a Heisenberg fluid. We intend to discuss this and other cases in a future paper.

\section{Appendix}
In curved space-time we used the covariant derivative of a spinor thus defined as
$$\nabla_{\mu} \, \Psi =\partial_{\mu} \, \Psi  - \mathbf{\Gamma}_{\mu}\, \Psi $$
in terms of the Fock-Ivanenko connection
$$\mathbf{\Gamma}_{\alpha} = - \, \frac{1}{8} \, \left(\gamma^{\mu} \, \gamma_{\mu \, ,\alpha} - \gamma_{\mu \, ,\alpha} \, \gamma^{\mu} - \Gamma^{\varrho}_{\alpha\mu}\, (\gamma^{\mu} \, \gamma_{\varrho} - \gamma_{\varrho} \, \gamma^{\mu}) \right)$$ in order to check the consistency required by the fundamental condition (eq. \ref{domingo6}).

\section{Acknowledgements} MN would like to thank CNPq for a grant.

\end{document}